\documentclass[prd,twocolumn,showpacs]{revtex4}
\usepackage{graphicx}
\usepackage{amsmath}
\usepackage{epsfig}

\begin{document}

\title{Pathways to testable leptogenesis}

\author{Pei-Hong Gu$^{1}_{}$}
\email{pgu@ictp.it}

\author{Utpal Sarkar$^{2}_{}$}
\email{utpal@prl.res.in}

\affiliation{ $^{1}_{}$The Abdus Salam International Centre for
Theoretical Physics, Strada Costiera 11, 34014 Trieste, Italy
\\
$^{2}_{}$Physical Research Laboratory, Ahmedabad 380009, India}

\begin{abstract}

In the conventional seesaw models of neutrino masses, leptogenesis
occurs at a very high scale. Three approaches have been discussed in
the literature to lower the scale of leptogenesis: mass degeneracy,
hierarchy of couplings and three-body decays. We advocate yet
another approach to a testable leptogenesis, whereby the decaying
particles could go out of equilibrium at an accessible scale due to
kinematics, although their couplings to the decay products are
larger for generating a desired CP asymmetry. We demonstrate this
new possibility for the testable leptogenesis in a two Higgs doublet
model where the neutrino masses originate from a one-loop diagram.

\end{abstract}


\maketitle

Among various baryogenesis mechanisms, which solve the puzzle of the
matter-antimatter asymmetry in the universe, the electroweak
baryogenesis is the most attractive one because of its testability.
However, it died in the standard model (SM) and only survives in the
minimal supersymmetric model (MSSM) with some rather restrictive
conditions or in the context of more complicated models.

At present, the leptogenesis
\cite{fy1986,lry1986,luty1992,mz1992,fps1995,fpsw1996,pilaftsis1997,ms1998,bpy2005}
scenario, where a lepton asymmetry is firstly generated and then is
partially converted to a baryon asymmetry by the sphalerons
\cite{krs1985}, is definitely the simplest baryogenesis theory to
test in the near future. Usually, the leptogenesis models also
explain the tiny but nonzero neutrino masses by the seesaw mechanism
\cite{minkowski1977,yanagida1979,grs1979,glashow1980,ms1980,sv1980}.
Since the tiny neutrino mass is naturally generated by a large
lepton number violating scale in the seesaw models, the decays of
heavier particles at the large lepton number violating scale also
generates the leptogenesis in these models.

Attempts have been made to lower the scale of leptogenesis so as to
make them testable in the next generation accelerators. Three
pathways to a testable leptogenesis model have been demonstrated in
some variants of the seesaw models through out-of-equilibrium
CP-violating decays of some TeV scale particles, which could be
detected at the LHC or ILC. These possibilities are: mass degeneracy
\cite{fps1995,pilaftsis1997} (the decaying particles are assumed to
be almost degenerate), hierarchy of couplings
\cite{hms2000,hambye2000} and three-body decays \cite{hambye2001}.
In the case of mass degeneracy, the decaying particles have tiny
couplings to guarantee their decay widths smaller than the expansion
rate of the universe, parametrized in terms of the Hubble constant
to satisfy the out-of-equilibrium condition. The decaying particles
have quasi-degenerate masses to resonantly enhance the CP asymmetry.
In the case of hierarchal couplings, the lighter decaying and
heavier virtual particles have smaller and larger couplings,
respectively, thus the smaller couplings determine the decay widths
while the large couplings dominate the CP asymmetry. In the case of
three-body decays, the decay widths of the lighter particles are
suppressed by the heavy masses of the virtual particles, whose
couplings are large enough to enhance the CP asymmetry. Note the
decaying particles should be gauge singlets of any low-energy gauge
symmetry, otherwise, the scattering processes originating from the
gauge interactions will considerably damp the produced lepton
asymmetry.

Obviously, the three pathways have a same essence that the
(effective) couplings of the decaying particles to the decay
products should be much smaller than unity so that the decay widths
could be smaller than the Hubble constant at a low scale. This
conclusion is based on an assumption that the decay products are
much lighter than the decaying particles and hence the decay widths
are only related to the masses of the decaying particles and the
couplings. However, it is not the only possibility
because the decay widths
could be kinematically suppressed even if the couplings are larger.

In this paper, we discuss the kinematical effect in the leptogenesis
scenario, which could alter the out-of-equilibrium condition considerably.
For demonstration, we consider a two Higgs doublet model
\cite{ma2006}, where the neutrino masses originate from a one-loop
diagram. Although the original model was proposed to include a
dark matter candidate, we relax this condition to allow a different
range of the parameters.
With the larger couplings and the proper masses of the
decaying particles and decay products, the sizable CP asymmetry and
the suppressed decay widths are simultaneously allowed to realize a
successful leptogenesis at an accessible scale.

We now introduce the quoted two Higgs doublet model \cite{ma2006},
where the SM is extended by three right-handed neutrinos $N_{R}^{}$
and one scalar doublet $\eta=(\eta^{0}_{},\eta^{-}_{})^{T}_{}$. The
new particles are odd under a $Z_{2}^{}$ discrete symmetry. As a
result, the following Yukawa couplings and Majorana mass term are
allowed,
\begin{eqnarray}
\label{lagrangian1} \mathcal{L}\supset -y\overline{\psi_L^{}}\eta
N_{R}^{}-\frac{1}{2}M_{N}^{}\overline{N_{R}^{c}}N_{R}^{}+\textrm{h.c.}\,,
\end{eqnarray}
where $\psi_{L}^{}=(\nu_{L}^{},\ell_{L}^{})^{T}_{}$ denotes the SM
left-handed leptons. For convenience and without loss of any
generality, we will choose the basis in which the Majorana mass
Matrix $M_{N}^{}$ is real and diagonal.

\begin{figure}
\vspace{6.0cm} \epsfig{file=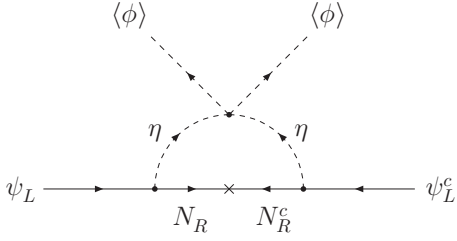, bbllx=5.3cm,
bblly=6.0cm, bburx=15.3cm, bbury=16cm, width=7cm, height=7cm,
angle=0, clip=0} \vspace{-9cm} \caption{\label{massgeneration} The
one-loop diagram for generating the radiative neutrino masses.}
\end{figure}

There are no Dirac masses between the left- and right-handed
neutrinos since the exact $Z_{2}^{}$ protects $\eta$ against any
vacuum expectation values ($vev$s). So, the neutrinos will remain
massless at tree level unless the quartic scalar interaction,
\begin{eqnarray}
\label{quartic} \mathcal{L}\supset
-\frac{1}{2}\lambda\left[\left(\phi^{\dagger}_{}\eta\right)^{2}_{}+\textrm{h.c.}\right]\,,
\end{eqnarray}
is introduced to realize the one-loop diagram as depicted in Fig.
\ref{massgeneration}. Here $\phi=(\phi^{0}_{},\phi^{-}_{})^{T}_{}$
is the SM Higgs with $\langle\phi\rangle\equiv v \simeq
174\,\textrm{GeV}$. For the purpose of demonstration, we will take
$\lambda$ to be positive in the following discussions and
calculations. The radiative neutrino masses can be explicitly
calculated \cite{ma2006},
\begin{eqnarray}
\label{mass} \left(m_{\nu}^{}\right)_{ij}^{}
&=&\frac{1}{16\pi^{2}_{}}\sum_{k=1}^{3}y^{}_{ik}
y^{}_{jk}M_{N_{k}^{}}^{}\nonumber\\
&&\times\left[\frac{M_{\eta^{0}_{R}}^{2}}{M_{\eta^{0}_{R}}^{2}-M_{N_{k}^{}}^{2}}
\ln
\left(\frac{M_{\eta^{0}_{R}}^{2}}{M_{N_{k}^{}}^{2}}\right)\right.\nonumber\\
&&\left.-\frac{M_{\eta^{0}_{I}}^{2}}{M_{\eta^{0}_{I}}^{2}-M_{N_{k}^{}}^{2}}
\ln
\left(\frac{M_{\eta^{0}_{I}}^{2}}{M_{N_{k}^{}}^{2}}\right)\right]\,,
\end{eqnarray}
where $\eta^{0}_{R}$ and $\eta^{0}_{I}$ are defined by
$\displaystyle{\eta^{0}_{}=\frac{1}{\sqrt{2}}\left(\eta^{0}_{R}+i\eta^{0}_{I}\right)}$.
For
$\displaystyle{M_{\eta^{0}_{R}}^{2}-M_{\eta^{0}_{I}}^{2}=2\lambda
v^{2}_{}\ll M_{\eta^{0}_{R,I}}^{2}}$ and
$M_{N_{k}^{}}^{2}>M_{\eta^{0}_{R,I}}^{2}$, we can obtain a simple
formula for the neutrino masses,
\begin{eqnarray}
\label{mass2} \left(m_{\nu}^{}\right)_{ij}^{}
=\frac{\mathcal{O}(\lambda)}{16\pi^{2}_{}}\sum_{k=1}^{3}\frac{y^{}_{ik}
y^{}_{jk}}{M_{N_{k}^{}}^{}}v^{2}_{}\,.
\end{eqnarray}
Therefore, the observed neutrino masses can be naturally explained,
for example, $m_{\nu}^{}\sim \mathcal{O}(0.01-0.1\,\textrm{eV})$ for
$\lambda=\mathcal{O}(10^{-4})$, $y\sim\mathcal{O}(10^{-3}_{})$ and
$M_{N_{i}^{}}^{}=\mathcal{O}(100\,\textrm{GeV}-10\,\textrm{TeV})$.

\begin{figure*}
\vspace{4.0cm} \epsfig{file=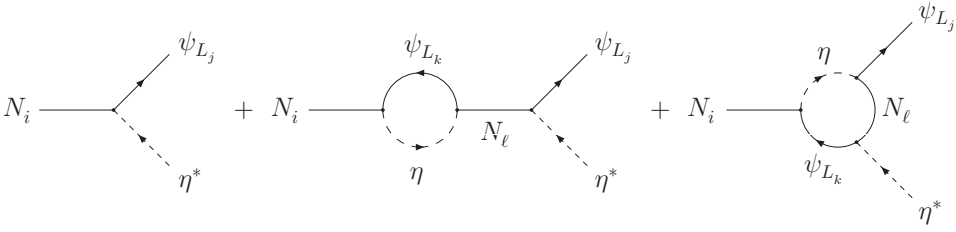, bbllx=5cm, bblly=6.0cm,
bburx=15cm, bbury=16cm, width=7cm, height=7cm, angle=0, clip=0}
\vspace{-7cm} \caption{\label{decay} The heavy Majorana neutrinos
decay into the SM left-handed leptons and the doublet scalar. Their
CP conjugate states have been omitted for simplicity.}
\end{figure*}

The heavy Majorana neutrinos $N=N_{R}^{}+N_{R}^{c}$ can decay into
the SM left-handed leptons and the doublet scalar. For simplicity,
only the decay channel of $N\rightarrow \psi_{L}^{}\eta^{\ast}_{}$
is shown in Fig. \ref{decay} and its CP conjugate state
$N\rightarrow \psi_{L}^{c}\eta$ is omitted.
At the tree level, the decay widths are given by
\begin{eqnarray}
\label{decaywidth} &&\Gamma\left(N_i^{}\rightarrow
\psi_L^{}+\eta^\ast_{}\right)=\Gamma\left(N_i^{}\rightarrow
\psi_L^{c}+\eta\right)\\
&=&\frac{1}{16\pi}\left(y^{\dagger}_{}y\right)_{ii}^{}\,M_{N_{i}^{}}^{}r^{2}_{N_{i}^{}}\,,
\end{eqnarray}
where the factor,
\begin{eqnarray}
\label{kinetics}
r_{N_{i}^{}}^{}=1-\frac{M_{\eta}^{2}}{M_{N_{i}^{}}^{2}}\,,
\end{eqnarray}
depends on the mass of the decay product in addition to that of the
decaying particle. For the increasing values of $M_{\eta}^{}$, we
get a smooth interpolation from $r_{N_{i}^{}}^{}\simeq 1~
(M_{\eta}^{2}\ll M_{N_{k}^{}}^{2})$ to $r_{N_{i}^{}}^{}\rightarrow 0
~ (M_{\eta}^{2}\rightarrow M_{N_{k}^{}}^{2})$. Usually the decay
products are assumed to be very light, and hence, this factor is
taken to be $1$. However, this factor can also be very small. We
shall consider the case when this factor is very small. In the
present model under consideration, we shall not provide any
explanation for the smallness of this parameter $r_{N_i}$. But for
purpose of demonstration let us give an explicit example, where this
parameter could be very small naturally.

Consider a supersymmetric model with superfields $N$ and $S$. An
Yukawa coupling $NNS$ would then allow the decay of the fermionic
component of $N$ to its lighter scalar component and the fermionic
component of $S$: $N_{f}^{} \rightarrow N_{s}^{} + S_{f}^{}$. If the
masses of these fields are $M_{N}^{} \sim 10^{7}_{}\,\textrm{GeV}$
and $M_{S}^{} \ll M_{N}^{}$, the mass splitting between $N_{f}^{}$
and $N_{s}^{}$ will be of the order of supersymmetry breaking scale
of $10^{3}_{}\,\textrm{GeV}$, so that this decay width will have a
suppression factor of $r_{N}^{} \sim 10^{-4}$. Although this example
can not be extrapolated to the present model, we shall assume a
similar small value for $r_{N_i}^{}$.

Coming back to the present model, the decays of the heavy Majorana
neutrinos can create a lepton asymmetry, if the Yukawa couplings $y$
provide the source of the CP violation. The lowest order non-trivial
asymmetry comes from the interference of the tree-level diagrams
with the one-loop diagrams. We calculate the CP asymmetry and find
it to be same as that in the canonical seesaw model,
\begin{eqnarray}
\label{cpasymmetry} \varepsilon_{N_i}^{}&\equiv & \frac{\Gamma
(N_i^{}\rightarrow \psi_L^{}+\eta^\ast_{})- \Gamma
(N_i^{}\rightarrow \psi_L^{c}+\eta)}{\Gamma (N_i^{}\rightarrow
\psi_L^{}+\eta^\ast_{})+ \Gamma (N_i^{}\rightarrow
\psi_L^{c}+\eta)}\nonumber\\
&\simeq&\frac{1}{8\pi}\frac{1}{\left(y^{\dagger}_{}y\right)_{ii}^{}}\sum_{j\neq
i}^{}\textrm{Im}\left[\left(y^{\dagger}_{}y\right)_{ij}^{2}\right]\nonumber\\
&&\times
\left[f\left(\frac{M_{N_{j}^{}}^{2}}{M_{N_{i}^{}}^{2}}\right)
+g\left(\frac{M_{N_{j}^{}}^{2}}{M_{N_{i}^{}}^{2}}\right)\right]\,,
\end{eqnarray}
which is free from the masses of the decay products. Here the
functions $f$ and $g$ are the contributions from the vertex and
self-energy corrections, respectively:
\begin{eqnarray}
\label{vertex}
f(x)&=&\sqrt{x}\left[1-(1+x)\ln\left(\frac{1+x}{x}\right)\right]\,,\\
\label{selfenergy} g(x)&=&\frac{\sqrt{x}}{1-x}\,.
\end{eqnarray}

For illustration, we consider the limiting case with
$M_{N_{1}^{}}^{}\ll M_{N_{2,3}^{}}^{}$ (a factor
$\displaystyle{M_{N_{2,3}^{}}^{}/M_{N_{1}^{}}^{}}\displaystyle$ of
$3-10$ is enough because the number density of $N_{2,3}^{}$ is
fastly Boltzmann suppressed at the temperature below its mass),
where the final lepton asymmetry should mainly come from the
contributions of the decays of $N_{1}^{}$. We can simplify the CP
asymmetry (\ref{cpasymmetry}) as
\begin{eqnarray}
\label{cpasymmetry2}
\varepsilon_{N_1}^{}&\simeq&-\frac{3}{16\pi}\frac{1}{\left(y^{\dagger}_{}y\right)_{11}^{}}\sum_{j=2,3
}^{}\textrm{Im}\left[\left(y^{\dagger}_{}y\right)_{1j}^{2}\right]\frac{M_{N_{1}^{}}^{}}{M_{N_{j}^{}}^{}}\,.
\end{eqnarray}
Similar to the DI bound \cite{di2002,bdp2003}, one can also deduce
an upper bound on the above CP asymmetry by inserting the neutrino
masses (\ref{mass2}) into the CP asymmetry (\ref{cpasymmetry2}),
\begin{eqnarray}
\label{cpasymmetry3} |\varepsilon_{N_1}^{}|&<&
\frac{3\pi}{\mathcal{O}(\lambda)}\frac{M_{N_{1}^{}}^{}m_{3}^{}}{v^{2}_{}}|\sin\delta|
\end{eqnarray}
with $m_3^{}$ and $\delta$ being the biggest eigenvalue of the
neutrino mass matrix and the CP phase, respectively. Here we have
assumed the neutrinos to be hierarchical \cite{sv2006}. Clearly, in
the present case, the DI bound is relaxed by a factor of
$16\pi^{2}_{}/\mathcal{O}(\lambda)$.

For effectively creating a lepton asymmetry in the thermal evolution
of the universe, the decays of $N_{1}^{}$ should satisfy the
condition of departure from equilibrium, which is described by
\begin{eqnarray}
\label{condition} \Gamma_{N_{1}^{}}^{}\lesssim
H(T)\left|_{T=M_{N_{1}^{}}^{}}^{}\right.\,.
\end{eqnarray}
where
\begin{eqnarray}
\label{decaywidth2} \Gamma_{N_{1}^{}}^{}&=&
\Gamma\left(N_{1}^{}\rightarrow
\psi_L^{}+\eta^\ast_{}\right)+\Gamma\left(N_{1}^{}\rightarrow
\psi_L^{c}+\eta\right)\nonumber\\
&=&\frac{1}{8\pi}\left(y^{\dagger}_{}y\right)_{11}^{}\,M_{N_{1}^{}}^{}r^{2}_{N_{1}^{}}
\end{eqnarray}
is the total decay width of $N_{1}^{}$ and
\begin{eqnarray}
\label{hubble}
H(T)&=&\left(\frac{8\pi^{3}_{}g_{\ast}^{}}{90}\right)^{\frac{1}{2}}_{}\frac{T^{2}_{}}{M_{\textrm{Pl}}^{}}
\end{eqnarray}
is the Hubble constant with the Planck mass
$M_{\textrm{Pl}}^{}\simeq 10^{19}_{}\,\textrm{GeV}$ and the
relativistic degrees of freedom $g_{\ast}^{}\simeq 100$
\cite{kt1990}.

It is straightforward to perform the out-of-equilibrium condition by
inserting (\ref{decaywidth2}) and (\ref{hubble}) to
(\ref{condition}),
\begin{eqnarray}
\left(y^\dagger
y\right)_{11}^{}&\lesssim&\left(\frac{256\cdot\pi^5_{}\cdot
g_\ast^{}}{45}\right)^{\frac{1}{2}}_{}\,\frac{M_{N_{1}^{}}^{}}{M_{\textrm{Pl}}^{}}\frac{1}{r^{2}_{N_{1}^{}}}
\end{eqnarray}
for $y\sim \mathcal{O}(10^{-3}_{})$,
$M_{N_{1}^{}}^{}=\mathcal{O}(100\,\textrm{GeV}-1\,\textrm{TeV})$ and
$r_{N_{1}^{}}^{}=\mathcal{O}(10^{-4}_{})$. So, by inputting
$\lambda=10^{-4}_{}$, $M_{N_{1}^{}}^{}=700\,\textrm{GeV}$,
$m_3^{}=0.07\,\textrm{eV}$ and $\sin\delta=-1$ to the CP asymmetry
(\ref{cpasymmetry3}), we derive $\varepsilon_{N_{1}^{}}^{}\simeq
-1.5\times 10^{-7}_{}$ and then obtain the final baryon asymmetry,
\begin{eqnarray}
\label{baryonasymmetry11}
\frac{n_B^{}}{s}&=&\frac{28}{79}\,\frac{n_{B-L}^{}}{s}=-\frac{28}{79}\,\frac{n_{L}^{}}{s}\nonumber\\
&\simeq&
-\frac{28}{79}\,\varepsilon_{N_{1}^{}}^{}\frac{n_{N_{1}^{}}^{eq}}{s}\left|_{T=M_{N_{1}^{}}^{}}^{}\right.\simeq
-\frac{1}{15}\frac{\varepsilon_{N_{1}^{}}^{}}{g_\ast^{}}\nonumber\\
&\simeq& 10^{-10}
\end{eqnarray}
as desired to explain the matter-antimatter asymmetry of the
universe.

In this leptogenesis scenario, all of the new particles are at an
accessible scale and then are expected to be detected at the forthcoming
experiments. Specifically, the $\eta$ scalars can be directly
produced in pairs by the SM gauge bosons $W^{\pm}_{}$, $Z$ or
$\gamma$. Once produced, $\eta^{\pm}_{}$ will decay into
$\eta_{R,I}^{0}$ and a virtual $W^{\pm}_{}$, which becomes a
quark-antiquark or lepton-antilepton pair. If $\eta^{0}_{R}$ is
heavier than $\eta^{0}_{I}$ for $\lambda>0$, the decay chain,
\begin{eqnarray}
\eta^{+}_{}\rightarrow \eta^{0}_{R}l^{+}_{}\nu \,, \quad
\textrm{then} \quad \eta^{0}_{R}\rightarrow
\eta^{0}_{I}l^{+}_{}l^{-}_{}\,,
\end{eqnarray}
has $3$ charged leptons and large missing energy, and can be
compared to the direct decay,
\begin{eqnarray}
\eta^{+}_{}\rightarrow \eta^{0}_{I}l^{+}_{}\nu\,,
\end{eqnarray}
to extract the masses of the respective particles.

\begin{figure}
\vspace{3.5cm} \epsfig{file=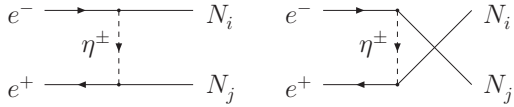, bbllx=2.3cm, bblly=6.0cm,
bburx=12.3cm, bbury=16cm, width=7cm, height=7cm, angle=0, clip=0}
\vspace{-8.5cm} \caption{\label{annihilation} The heavy Majorana
neutrinos $N_{1,2,3}^{}$ are produced in pairs by the
$e^{+}_{}e^{-}_{}$ annihilation through the $\eta^{\pm}_{}$
exchanges,}
\end{figure}

As for the heavy Majorana neutrinos $N_{1,2,3}^{}$, they can be
produced in pairs by the $e^{+}_{}e^{-}_{}$ annihilation through the
$\eta^{\pm}_{}$ exchanges as shown in Fig. \ref{annihilation}, which
could be significant for the larger Yukawa couplings. We give the
cross sections as below,
\begin{eqnarray}
\sigma(e^{+}_{}e^{-}_{}\rightarrow N_{i}^{}N_{i}^{})&=&\frac{1}{32\pi}|y_{ei}^{}|^{4}_{}\left(F_{ii}^{}+H_{ii}^{}\right)\,,\\
\sigma(e^{+}_{}e^{-}_{}\rightarrow
N_{i}^{}N_{j}^{})&=&\frac{1}{32\pi}|y_{ei}^{}|^{2}_{}|y_{ej}^{}|^{2}_{}F_{ij}^{}\,,
\end{eqnarray}
where
\begin{eqnarray}
F_{ij}^{} &=&\frac{1}{s\left(s-4m_{e}^{2}\right)}
\left\{\left(2M_{\eta^{\pm}_{}}^{2}-2m_{e}^{2}-M_{N_{i}^{}}^{2}-M_{N_{j}^{}}^{2}\right)\right.\nonumber\\
&&\times
\ln\left|\frac{t_{0}^{}-M_{\eta^{\pm}_{}}^{2}}{t_{1}^{}-M_{\eta^{\pm}_{}}^{2}}\right|\nonumber\\
&&+\left[\frac{\left(M_{\eta^{\pm}_{}}^{2}-M_{N_{i}^{}}^{2}-m_{e}^{2}\right)
\left(M_{\eta^{\pm}_{}}^{2}-M_{N_{j}^{}}^{2}-m_{e}^{2}\right)}
{\left(t_{0}^{}-M_{\eta^{\pm}_{}}^{2}\right)\left(t_{1}^{}-M_{\eta^{\pm}_{}}^{2}\right)}\right.\nonumber\\
&&\left.\left.+1\right]\left(t_{0}^{}-t_{1}^{}\right)\right\}\,,\\
\end{eqnarray}
\begin{eqnarray}
H_{ii}^{}&=&\frac{M_{N_{i}^{}}^{2}\left(s-2m_{e}^{2}\right)}{s\left(s-4m_{e}^{2}\right)\left(s-2m_{e}^{2}-2M_{N_{i}^{}}^{2}+2M_{\eta^{\pm}_{}}^{2}\right)}\nonumber\\
&&\times\ln\left|\frac{t_{0}^{}-M_{\eta^{\pm}_{}}^{2}}{t_{1}^{}-M_{\eta^{\pm}_{}}^{2}}\right|\,,
\end{eqnarray}
with
\begin{eqnarray}
t_{0}^{}(t_{1}^{})&=&
-\frac{1}{4}\left\{\left(s-4m_{e}^{2}\right)^{\frac{1}{2}}_{}\mp
\left[\frac{s-\left(M_{N_{i}^{}}^{}-M_{N_{j}^{}}^{}\right)^{2}_{}}{s}\right]^{\frac{1}{2}}_{}\right.\nonumber\\
&&\left.\times
\left[s-\left(M_{N_{i}^{}}^{}+M_{N_{j}^{}}^{}\right)^{2}_{}\right]^{\frac{1}{2}}_{}\right\}^{2}_{}\,.
\end{eqnarray}

In this paper, we point out the kinematical approach to the testable
leptogenesis. The kinematical effect allows the (effective)
couplings of the decaying particles to the decay products to be
larger for the sizable CP asymmetry, meanwhile, allows the decay
widths to be smaller for the out-of-equilibrium condition at an
accessible scale. We demonstrate this scenario in a two Higgs
doublet model where the Majorana neutrino masses originate from a
one-loop diagram. We can also apply the kinematical effect to the
model where the neutrinos are Dirac particles and their tiny masses
are radiatively generated \cite{gs2007}. These leptogenesis models
can be verified in the near future.

\end{document}